\documentclass[12pt]{article}

\usepackage{amsmath}

\usepackage{amsfonts}

\usepackage{amssymb}

\usepackage[latin1]{inputenc}

\usepackage{graphics}

\usepackage{graphicx}

\usepackage{epsfig}

\usepackage{amsthm}

\usepackage{lineno}

\usepackage{mathrsfs}

\usepackage{latexsym}

\newcommand{\h}{\hspace{.5cm}}

\newenvironment{resumo}{\begin{quotation}\small\em}{\end{quotation}}
\date{}
\title{Neumann boundary conditions inhibiting the SSB in the Coleman-Weinberg mechanism}
\author{{\bf F. N. Fagundes\footnote{Present  address: Universidade Federal de Viçosa, Brasil.}, R. O. Francisco, B. B. Dilem and J. A. Nogueira\footnote{Electronic address: nogueira@cce.ufes.br}}\\
{\it Universidade Federal do Esp\'{\i}rito Santo, Brasil}}
\begin{document}

\maketitle

\begin{abstract}
\begin{resumo}
In this work we show that homogeneous Neumann boundary conditions inhibit the Coleman-Weinberg mechanism for spontaneous symmetry breaking in the scalar electrodynamics if the length of the finite region is small enough ($a = e^{2}M^{-1}_{\phi}$, where $M_{\phi}$ is the mass of the scalar field generated by the Coleman-Weinberg mechanism).
\end{resumo}
\end{abstract}


\pagenumbering{arabic}
\section{Introduction}
\label{secint}

\h The spontaneous symmetry breaking plays an essential role in the Higgs mechanism for the mass generation of the fundamental particles. 
In the standard model the spontaneous symmetry breaking is induced by the Higgs potential through an imaginary mass term inserted by hand. 
The arbitrariness of the Higgs potential is one of the weaknesses of the Higgs mechanism, since many of the physical parameters depend on 
the precise form of the Higgs potential \cite{kane,kaku}.
In an alternative approach of Coleman-Weinberg \cite{cw}, the spontaneous symmetry breaking is induced by 1-loop radiative corrections, 
rather than being inserted by hand. In the Coleman-Weinberg approach of the scalar electrodynamics the spontaneous symmetry breaking occurs when the coupling constant $\lambda$ of the self-interaction of the scalar field is of the same order of $e^{4}$, the constant of minimal coupling, turning terms from radiative corrections of classical order. 
On the other hand, extraordinary effects arise if we consider the Quantum Field Theory not in the infinite space, as usual, but in the space restricted by some boundary surfaces with the respective boundary conditions satisfied by the fields or in spaces with non-Euclidean topology. Of particular interest is the restorer symmetry \cite{tomsma, tomsma2, fulling, sissakian, denardo, elizaldessb, kennedy, ze}. The scalar electrodynamics in a flat space-time with the topoloy $S^{1} \times R^{3}$ has been examined by T. Yoshimura \cite{yoshimura} and D. Toms \cite{toms}.
Boundary conditions introduce a new parameter, the size of the finite region, which changes the radiative corrections. So, it is of interest to consider the possibility that the spontaneous symmetry breaking in the Coleman-Weinberg mechanism may be inhibited by the boundary conditions. 
The purpose of this work is to find out whether Neumann boundary conditions are able to inhibit the Coleman-Weinberg mechanism. To this end we consider the scalar and vector fields of the massless scalar electrodynamics satisfying homogeneous Neumann boundary conditions on two infinite parallel surfaces separated by some small distance $a$. We show that there is a maximum length, named 
{\it critical length} $a_{c}$, so that the spontaneous symmetry breaking does not occur.
Although we use the scalar electrodynamics and Neumann boundary conditions as a first example of inspiration to achieve our final aim of investigating the possibility that boundary conditions inhibit the Coleman-Weinberg mechanism for the spontaneous symmetry breaking, the scalar electrodynamics has been used as a prototype for quantum chromodynamics, in which Neumann boundary conditions are present \cite{appelquist,blaizot}. Moreover, the scalar electrodynamics has been also considered, at the beginning, as the effective theory of matter and forces in the brane worlds \cite{kaviani}. So, this example may indeed be used for later investigation with a realistic model.
Since homogeneous Neumann boundary conditions allow constant value of the field, we use the effective potential for checking  if $\langle \phi \rangle = 0$ is the vacuum (minimum) of the theory.

The outline of our paper is as follows: in Section \ref{secepse} we derive the effective potential for scalar electrodynamics with the scalar and vector fields satisfying homogenous Neumann boundary conditions on two infinite parallel plane surfaces separated by some small distance $a$. In Section \ref{secmsf} we calculate the first derivative of the effective potential and we examine the second derivative at $\left\langle \phi \right\rangle = 0$ in order to show that if the length of the finite region is small enough, the zero field solution is stable (vacuum), in consequence the spontaneous symmetry breaking is inhibited. In Section \ref{seccon} we present our conclusions.

\section{Effective Potential for the Scalar Electrodynamics}
\label{secepse}

\h Let us consider the theory of a massless, quartically 
self-interacting complex scalar field $\phi(x)$ minimally coupled to the electromagnetic field.
The lagrangian density for this theory is \cite{ryder}
\begin{equation}
\label{l1}
{\cal L} = - \frac{1}{4}F_{\mu\nu}F^{\mu\nu} +
 D_{\mu }\phi^{*}D^{\mu }\phi - 
\frac{\lambda }{6}\left(\phi^{*}\phi\right)^{2} - \frac{1}{2\xi}\left(\partial^{\mu}A_{\mu}\right)^{2}
 - \bar{\eta}\partial^{\mu}\partial_{\mu}\eta,
\end{equation}
where $D_{\mu}$ is the covariant derivate, given by
\begin{equation}
D_{\mu} = \partial _{\mu } + ieA_{\mu},
\end{equation}
necessary to keep the lagrangian invariant under transformation
 of the U(1) group and it is answerable for the minimally coupled. The fourth term is the gauge fixing  and last 
one is the Faddev-Popov term in which $\eta$ and $\bar{\eta}$ denote the ghost fields.

The contribution to the effective potential can be somewhat simplified 
writing the complex field in terms of two real fields $\phi_{1}$ and 
$\phi_{2}$. Putting
\begin{equation}\phi = \frac{1}{\sqrt{2}}\left( \phi_{1} + 
i\phi_{2} \right),
\end{equation}
and performing the usual Wick rotations (i.e. $x_{0} \rightarrow -ix_{4}$ and $A_{0} \rightarrow iA_{4}$), the 
lagrangian (\ref{l1}), in euclidian space-time, becomes
$$
{\cal L} = -\frac{1}{2}\left(\partial \phi_{a}\right)^{2} 
- \frac{1}{4}F_{\mu\nu}F^{\mu\nu} - \frac{1}{2\xi}\left(\partial^{\mu}A_{\mu}\right)^{2}
$$
\begin{equation}
\label{l2}
- \frac{1}{2}e^{2}A^{\mu}A_{\mu}\phi_{a}\phi_{a} - e\epsilon_{ab}\phi_{a}A^{\mu}\partial_{\mu}\phi_{b}
- \frac{\lambda }{4!}\left(\phi_{a}\phi_{b}\right)^{2} + \bar{\eta}\Box\eta,
\end{equation}
where $a, b = 1, 2$ and $\epsilon_{ab}$ is an antisymmetric tensor with $\epsilon_{12} = -\epsilon_{21} = 1$.

The one-loop effective potential\footnote{We keep $\hbar$ to mark the quantum corrections, but we set $\hbar = c = 1$ everywhere else.} is given by  \cite{toms, jackiw, ze2}
$$
V_{ef}^{(1)}(\phi_{c}) = \frac{\hbar}{2\Omega}\ln\left[\det\left(k^{2}+\frac{\lambda}{2}\phi_{c}^{2}\right)\right]+ \frac{3\hbar}{2\Omega}\ln[\det(k^{2}+e^{2}\phi_{c}^{2})]
$$
$$
+ \frac{\hbar}{2\Omega}\ln[\det(k^{2}+M_{+}^{2}\phi_{c}^{2})]+\frac{\hbar}{2\Omega}\ln[\det(k^{2}+M_{-}^{2}\phi_{c}^{2})]
$$
\begin{equation}
-2\frac{\hbar}{2\Omega}\ln[\det(k^{2}+m^{2})] ,
\end{equation}
where $\phi_{c}^{2} = \phi_{1c}^{2} + \phi_{2c}^{2}$ is the classical field, $\Omega$ is the four dimensional space-time volume  and
\begin{equation}
M_{\pm}^{2}=\frac{\lambda}{12}\pm\sqrt{\frac{\lambda^{2}}{144}-\frac{\xi\lambda e^{2}}{6}} .
\end{equation}
We have introduced $m^{2}$ in the last term, which is due to the ghost, and then we will take $m^{2} \rightarrow 0$ at the end in order to remove the infrared divergence.

Now we consider the fields satisfying homogeneous Neumann boundary conditions on two infinite parallel plane surfaces 
separated by some small distance $a$,
\begin{equation}
\label{nc}
\left. \frac{\partial \varphi}{\partial z}\right| _{z=0} = \left. \frac{\partial \varphi}{\partial z}\right| _{z=a} = 0,
\end{equation}
where $\varphi$ represents the fields $\phi_{a}$, $A^{\mu}$, $\eta$ and $\bar{\eta}$.
The boundary conditions (\ref{nc}) become the momentum perpendicular to surfaces ($z=0$ and $z=a$) discrete: $k_{z} = \frac{n \pi}{a}$, where $n = 0, 1, 2, 3, \dots$, whereas others remain continuous. So, the term in order $\hbar$ of the effective potential becomes 
$$
V^{(1)}_{ef}\left(\phi_{c}\right) =  
\frac{1}{2}\frac{\hbar }{\Omega }\ln \det \left[ \bar{k}^{2} + \frac{n^{2}\pi^{2}}{a^{2}} + \beta^{2}\phi_{c}^{2} \right] 
+ \frac{3}{2}\frac{\hbar }{\Omega }\ln \det \left[\bar{k}^{2} + \frac{n^{2}\pi^{2}}{a^{2}} + e^{2}\phi_{c}^{2} \right]
$$
$$
+ \frac{1}{2}\frac{\hbar }{\Omega }\ln \det \left[ \bar{k}^{2} + \frac{n^{2}\pi^{2}}{a^{2}} + M^{2}_{+}\phi_{c}^{2} \right] 
+ \frac{1}{2}\frac{\hbar }{\Omega }\ln \det \left[\bar{k}^{2} + \frac{n^{2}\pi^{2}}{a^{2}} + M^{2}_{-}\phi_{c}^{2} \right]
$$
\begin{equation}
\label{v1}
- 2\frac{1}{2}\frac{\hbar }{\Omega }\ln \det \left[\bar{k}^{2} + \frac{n^{2}\pi^{2}}{a^{2}} + m^{2} \right] ,
\end{equation}
where $\lambda = 2\beta^{2}$ and $k^{2} = \bar{k}^{2} + k_{z}^{2}$.

Now, using the procedure of the zeta function regularization \cite{hawking, elizalde, ramond} we obtain
$$
V_{ef}\left( \phi_{c} \right) = \frac{\beta^{2}}{12}\phi_{c}^{4} - \frac{\hbar\pi^{2}}{720a^{4}}
$$
$$
+  \frac{\hbar}{64\pi^{2}}\beta^{4}\phi_{c}^{4}\left[ \ln\left(\frac{\beta^{2} \phi_{c}^{2}}{\mu^{2}}\right) - \frac{3}{2} \right]
- \frac{\hbar \beta^{3}\phi_{c}^{3}}{24a\pi}  
- \frac{\hbar}{8\pi ^{2}}\beta^{4}\phi_{c}^{4} \sum_{n=1}^{\infty }\frac{K_{2}\left(  2n\beta a \phi_{c} \right) } {\left( n\beta a\phi_{c} \right)^{2}}
$$
$$
+  \frac{3\hbar}{64\pi^{2}}e^{4}\phi_{c}^{4}\left[ \ln\left( \frac{e^{2} \phi_{c}^{2}}{\mu^{2}} \right) - \frac{3}{2} \right]
- \frac{\hbar e^{3}\phi_{c}^{3}}{8a\pi}  
- \frac{3\hbar}{8\pi ^{2}}e^{4}\phi_{c}^{4} \sum_{n=1}^{\infty }\frac{K_{2}\left( 2nea\phi_{c} \right) } {\left( nea\phi_{c} \right) ^{2}} 
$$
$$
+  \frac{\hbar}{64\pi^{2}}M_{+}^{4}\phi_{c}^{4}\left[ \ln\left(\frac{M_{+}^{2} \phi_{c}^{2}}{\mu^{2}}\right) - \frac{3}{2} \right]
- \frac{\hbar M_{+}^{3}\phi_{c}^{3}}{24a\pi}  
- \frac{\hbar}{8\pi ^{2}}M_{+}^{4}\phi_{c}^{4} \sum_{n=1}^{\infty }\frac{K_{2}\left( 2nM_{+} a\phi_{c} \right)}{\left( nM_{+} a\phi_{c}\right)^{2}} 
$$
\begin{equation}
\label{l3}
+  \frac{\hbar}{64\pi^{2}}M_{-}^{4}\phi_{c}^{4}\left[ \ln\left(\frac{M_{-}^{2} \phi_{c}^{2}}{\mu^{2}}\right) - \frac{3}{2} \right]
- \frac{\hbar M_{-}^{3}\phi_{c}^{3}}{24a\pi}  
- \frac{\hbar}{8\pi ^{2}}M_{-}^{4}\phi_{c}^{4} \sum_{n=1}^{\infty }\frac{K_{2}\left( 2nM_{-} a\phi_{c} \right)}{\left( nM_{-} a\phi_{c}\right)^{2}},
\end{equation}
where $K_{\nu }(x)$ are modified Bessel functions. We have introduced an unknown scale parameter $\mu$, with dimensions of (length)$^{-1}$ or mass in order to keep the zeta function dimensionless for all s.

\section{$\left\langle \phi_{c} \right\rangle = 0 $ as Minimum of the $V_{ef}\left( \phi_{c} \right)$}
\label{secmsf}

\h Differentiating Eq. (\ref{l3}) we have
$$
\frac{dV_{ef}}{d \phi_{c}} =  \frac{\beta^{2}}{3}\phi_{c}^{3}
$$
$$
+  \frac{\hbar}{16\pi^{2}}\beta^{4}\phi_{c}^{3}\left[ \ln\left(\frac{\beta^{2} \phi_{c}^{2}}{\mu^{2}}\right) - 1 \right]
- \frac{\hbar \beta^{3}\phi_{c}^{2}}{8 \pi a}  
+ \frac{\hbar}{4\pi ^{2} a}\beta^{3}\phi_{c}^{2} \sum_{n=1}^{\infty }\frac{K_{1}\left(  2n \beta a \phi_{c} \right) } {n}
$$
$$
+  \frac{3 \hbar}{16\pi^{2}}e^{4}\phi_{c}^{3}\left[ \ln\left( \frac{e^{2} \phi_{c}^{2}}{\mu^{2}} \right) - 1 \right]
- \frac{3 \hbar e^{3}\phi_{c}^{2}}{8\pi a}  
+ \frac{3\hbar}{4\pi ^{2} a}e^{3}\phi_{c}^{2} \sum_{n=1}^{\infty }\frac{K_{1}\left( 2nea\phi_{c} \right) } {n} 
$$
$$
+  \frac{\hbar}{16 \pi^{2}}M_{+}^{4}\phi_{c}^{3}\left[ \ln\left(\frac{M_{+}^{2} \phi_{c}^{2}}{\mu^{2}}\right) - 1 \right]
- \frac{\hbar M_{+}^{3}\phi_{c}^{2}}{8 \pi a}  
- \frac{\hbar}{4\pi ^{2} a}M_{+}^{3}\phi_{c}^{2} \sum_{n=1}^{\infty }\frac{K_{1}\left( 2nM_{+} a\phi_{c} \right)}{n} 
$$
\begin{equation}
\label{d1}
+  \frac{\hbar}{16 \pi^{2}}M_{-}^{4}\phi_{c}^{3}\left[ \ln\left(\frac{M_{-}^{2} \phi_{c}^{2}}{\mu^{2}}\right) - 1 \right]
- \frac{\hbar M_{-}^{3}\phi_{c}^{2}}{8 \pi a}  
- \frac{\hbar}{4\pi ^{2} a}M_{-}^{3}\phi_{c}^{2} \sum_{n=1}^{\infty }\frac{K_{1}\left( 2nM_{-} a\phi_{c} \right)}{n}.
\end{equation}

It is easy to see that $\left\langle \phi_{c} \right\rangle = 0 $ is a solution to $\frac{dV_{ef}}{d \phi_{c}} = 0$. In order to see if there is any spontaneous symmetry breaking  we take into account the second derivative of the effective potential at  $\left\langle \phi_{c} \right\rangle = 0 $,
\begin{equation}
\label{d2}
\left. \frac{d^{2}V_{ef}}{d\phi_{c} ^{2}}\right| _{\phi_{c}=\left\langle\phi_{c}\right\rangle = 0 }  =  \frac{\hbar}{48a^{2}} \left( \frac{4 \beta^{2}}{3} + 3 e^{2} \right).
\end{equation}

Now, we make the assumption that $\beta$ is of order $e^{2}$, since we wish to investigate if the homogeneous Neumann boundary conditions may inhibit the Coleman-Weinberg mechanism for spontaneous symmetry breaking. Then the term of order $\beta^{2}$ in Eq. (\ref{d2}) is negligible when compared with the term of order $e^{2}$ and we can drop it, so
\begin{equation}
\label{d2a}
\left. \frac{d^{2}V_{ef}}{d\phi_{c} ^{2}}\right| _{\phi_{c}=\left\langle\phi_{c}\right\rangle = 0 }  =  \frac{\hbar e^{2}}{16a^{2}} = m_{\phi}^{2}.
\end{equation}

When $a \rightarrow \infty$ we may set the length scale as $(M_{\phi})^{-1}$, where $M_{\phi}$ is the mass of the scalar field acquired as a result of Coleman-Weinberg mechanism, in which $ M_{\phi} \approx e^{2} \left\langle \phi \right\rangle$. If $a \approx (M_{\phi})^{-1}$ then  $m_{\phi}^{2}\phi^{2} \approx e^{6} \left\langle \phi \right\rangle^{4}$ which is of higher order than the terms in $\phi^{4}$ of the effective potential. So, it can not modify the minimum and spontaneous symmetry breaking takes place (see Fig. 2).  If $a \approx e(M_{\phi})^{-1}$ then  $m_{\phi}^{2}\phi^{2} \approx e^{4} \left\langle \phi \right\rangle^{4}$ which is of the same order as the terms in $\phi^{4}$ of the effective potential. So, we can not disregard the term in $\phi^{2}$ arising from imposing the boundary conditions. However, the boundary conditions do not inhibit the spontaneous symmetry breaking, although they modify the minimum (see Fig. 3). If $a \approx e^{2}(M_{\phi})^{-1}$ then  $m_{\phi}^{2}\phi^{2} \approx e^{-2} \left\langle \phi \right\rangle^{4}$ which is of lower order than the terms in $\phi^{4}$ of the effective potential, therefore the terms in $\phi^{4}$ of the effective potential are negligible when compared with it. Consequently the Neumann boundary conditions inhibit the spontaneous symmetry breaking (see Fig. 4). 

In order to illustrate our previous statements, we have plotted the effective potential given by Eq. (\ref{l3}) as function of $\phi_{c}$ for some values of $a$ when $\beta=\frac{3e^{2}}{4\pi}$. For the sake of simplicity, we have chosen the Landau gauge ($\xi = 0$) and, since $\mu$ is arbitrary parameter, we have chosen $\mu = 100$ GeV.

In Fig. 1 is shown the effective potential when $a \rightarrow \infty$. That is the Coleman-Weinberg result. As it is clearly seen from Fig. 1 the minimum occurs at $ \left\langle \phi_{c} \right\rangle = \frac{\mu}{e}$.
\begin{figure}
\centering
\includegraphics[scale=0.8]{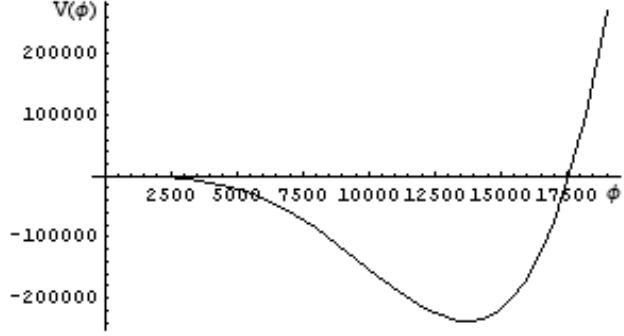} 
\caption{$V_{ef}(\phi_{c})$ X $\phi_{c}$ for $a \rightarrow \infty$. SSB induced by the Coleman-Weinberg mechanism.}
\end{figure}

In Fig. 2 is shown the effective potential when $a = \frac{1}{M_{\phi}}$, that is, $a$ is big. As we expected, the minimum occurs at $ \left\langle \phi_{c} \right\rangle = \frac{\mu}{e}$ again. This is because the effects of the corrections of the interactions of the vacuum fluctuations with the boundary are much smaller than the corrections of the self-interaction of the vacuum fluctuations, so they are negligible.
\begin{figure}
\centering
\includegraphics[scale=0.8]{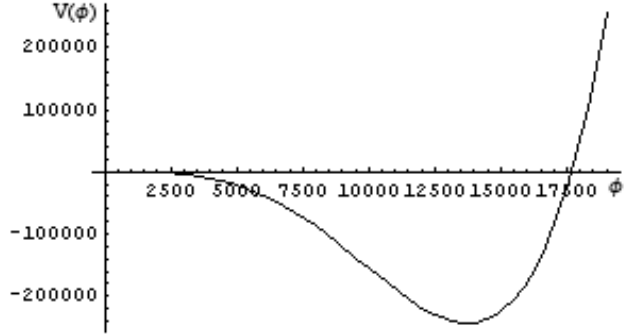} 
\caption{$V_{ef}(\phi_{c})$ X $\phi_{c}$  for $a \approx (M_{\phi})^{-1}$. The length of the finite region $a$ is big and the corrections due to the boundary conditions are negligible.}
\end{figure}

In Fig. 3 is shown the effective potential when $a = \frac{e}{M_{\phi}}$. Although $a$ is not big, it is not small enough for inhibiting the spontaneous symmetry breaking. However, note that the minimum is shifted away from $ \phi_{c} = \frac{\mu}{e}$. This is because the effects of the corrections of the interactions of the vacuum fluctuations with the boundary are of same order of the corrections of the self-interaction of the vacuum fluctuations, so they are not negligible.
\begin{figure}
\centering
\includegraphics[scale=0.8]{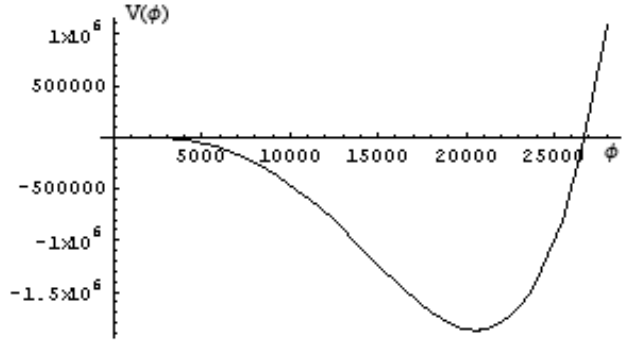} 
\caption{$V_{ef}(\phi_{c})$ X $\phi_{c}$ for $a \approx e(M_{\phi})^{-1}$. Although the length of the finite region $a$ is not big, it is not small enough for inhibiting the spontaneous symmetry breaking. However note that the minimum is shifted away from $ \phi_{c} = \frac{\mu}{e}$.}
\end{figure}

At last, in Fig. 4 is shown the effective potential when $a = \frac{e^{2}}{M_{\phi}}$. As we have said, now the minimum occurs at $ \left\langle \phi_{c} \right\rangle = 0$ and the spontaneous symmetry breaking does not take place. In this case the effects of the corrections of the self-interaction of the vacuum fluctuations are much smaller than the corrections of the interactions of the vacuum fluctuations with the boundary, so they are negligible. This way, we can say the corrections of the interactions of the vacuum fluctuations with the boundary inhibit the spontaneous symmetry breaking.
\begin{figure}
\centering
\includegraphics[scale=0.8]{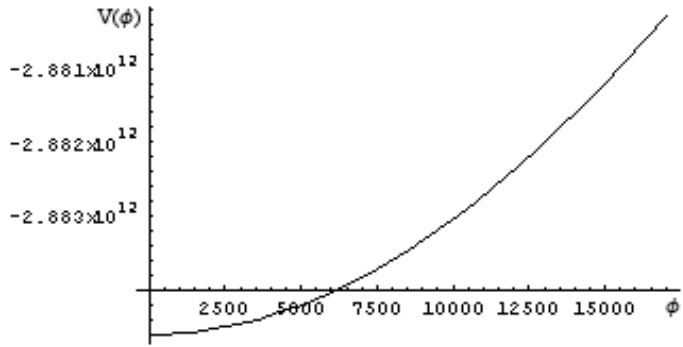} 
\caption{$V_{ef}(\phi_{c})$ X $\phi_{c}$  for $a \approx e^{2}(M_{\phi})^{-1}$.The length of finite region $a$ is small enough and SSB is inhibited.}
\end{figure}

We have worked in the one-loop approximation. So, it is natural to wonder whether the inhibition of the spontaneous symmetry breaking will occur when higher-order terms are included in the effective potential. Although an extending of the calculation to higher-order would be quite laborious, we can carry a rough estimate out. A dimensional analysis shows that the terms arising from imposing the boundary conditions are of the type $\phi^{n} a ^{n-4}$ ($n = 0, 1, 2 \dots$), for any order of $\hbar$. Since corrections of higher-order in $\hbar$ only increase the order of the coupling constants in the coefficients of the terms in powers of $\phi$, these terms will be negligible when compared with their respective terms in order of $\hbar$ (if $\beta$ is of order $e^{2}$). What makes our result reliable for all orders in $\hbar$.

It is clear from Eq. (\ref{d2}) that our result is gauge-invariant.

\section{Conclusion}
\label{seccon}

\h In this paper we have studied how the homogeneous Neumann boundary conditions on the fields of the scalar electrodynamics affect the Coleman-Weinberg mechanism for the spontaneous symmetry breaking. We have found if the length of the finite region, $a$, is small enough in such a way that the terms in $\phi^{4}$ of the effective potential are negligible when compared with the term in $\phi^{2}$ arising from imposing the Neumann boundary conditions, the spontaneous symmetry breaking due to the Coleman-Weinberg mechanism is inhibited.  It follows that there is a typical length scale given by the length  $a_{c}  = e^{2}(M_{\phi})^{-1}$ so that for $a < a_{c}$ spontaneous symmetry breaking does not occur.
Although our result is guaranteed only in the one-loop approximation, a rough estimate makes it reliable to higher-order. It is also worth noting the fact our result is gauge-invariant (at least in the order that we have worked).
The result obtained encourages the interest of continuing the discussion about boundary conditions to inhibit the spontaneous symmetry breaking with realistic (although more difficult) situation.
\\
\\
\\
{\bf Acknowledgments}

\h  This work was supported in part by CAPES, CNPq and PETROBRAS.\\



\end{document}